\documentclass[manuscript]{aastex}

\begin{document}


\title{SIGGMA: A Survey of Ionized Gas in the Galaxy, Made with the Arecibo Telescope}


\author{B. Liu\altaffilmark{1,2}}
\affil{National Astronomical Observatories, Chinese Academy of Sciences, Beijing 100012, China}


\author{T. Mclntyre\altaffilmark{}}
\affil{University of New Mexico, Albuquerque, NM 87131, USA}
\author{Y. Terzian}
\affil{Cornell University, Ithaca, NY 14853, USA}
\author{R. Minchin}
\affil{Arecibo Observatory, HC03 Box 53995, Arecibo 00612, Puerto Rico}
\author{L. Anderson}
\affil{West Virginia University, Morgantown, WV 26506, USA}
\author{E. Churchwell}
\affil{University of Wisconsin-Madison, Madison, WI 53706, USA}
\author{M. Lebron\altaffilmark{}}
\affil{University of Puerto Rico,PO Box 23323, 00931-3323, San Juan, Puerto Rico}

\and
\author{D. Anish Roshi\altaffilmark{}}
\affil{National Radio Astronomy Observatory, Green Bank \& Charlottesville, VA 22903, USA}


\altaffiltext{1}{State Key Radio Astronomy Laboratory of China}
\altaffiltext{2}{Graduate University of Chinese Academy of Sciences}


\begin{abstract}
A Survey of Ionized Gas in the Galaxy, made with the Arecibo telescope (SIGGMA) uses the Arecibo L-band Feed Array (ALFA) to fully sample the Galactic plane ($30 \degr \leq l \leq 75 \degr$ and $ -2 \degr \leq b \leq 2 \degr; 175 \degr \leq l \leq 207 \degr $ and $ -2 \degr \leq b \leq 1 \degr$) observable with the telescope in radio recombination lines (RRLs).
Processed data sets are being produced in the form of data cubes of $2 \degr$ (along $l$)$ \times 4 \degr$(along $b$)$ \times 151$ (number of channels), archived and made public. 
The 151 channels cover a velocity range of 600 km s$^{-1}$ and the velocity resolution of the survey changes from 4.2 km s$^{-1}$ to 5.1 km s$^{-1}$ from the lowest frequency channel to the highest frequency channel, respectively.
RRL maps with $3.4 \arcmin$ resolution and line flux density sensitivity of $\sim 0.5$~mJy will enable us to identify new HII regions, measure their electron temperatures, study the physics of photodissociation regions (PDRs) with carbon RRLs, and investigate the origin of the extended low density medium (ELDM).
Twelve Hn$\alpha$ lines fall within the 300 MHz bandpass of ALFA; they are resampled to a common velocity resolution to improve the signal-to-noise ratio (SN) by a factor of 3 or more and preserve the line width.
SIGGMA will produce the most sensitive fully sampled RRL survey to date.
Here we discuss the observing and data reduction techniques in detail.
A test observation toward the HII region complex S255/S257 has detected 
Hn$\alpha$ and Cn$\alpha$ lines with SN $>$ 10.
\end{abstract}


\keywords{ Galaxy: Structure - HII regions -radio lines: ISM -surveys
}


\section{Introduction}
\subsection{Background}
Recombination lines are emitted when electrons in ionized gas recombine with atomic nuclei to an excited state and cascade down in energy levels, $n$.
The most probable transitions are for change in energy level $\Delta n=1$ and are called `$\alpha$' lines. Transitions with $\Delta n = 2 $ are known as `$\beta$' lines and so on.
For hydrogen, $\alpha$ lines with $n \gtrsim 40 $ are in the radio regime ($\lambda \gtrsim 3$\,mm), and are termed radio recombination lines (RRLs)\citep[see][for a full account of the generation of RRLs]{gord2002}.

There are three main sources of RRL emission in our Galaxy: HII regions, diffuse ionized gas, and photodissociation regions (PDRs). HII regions are zones of plasma surrounding massive young stars.
Astrophysical RRLs were first detected toward the Omega HII region in 1964.
These observations were reported by \citet{drav1966} and \citet{soro1966}, and shortly afterward, were followed by high SN RRL detections toward Orion and M17 \citep{hogl1965}.
There have been many subsequent RRL surveys of HII regions \citep[e.g.][]{reif1970,wils1970,wils1980,lock1989,ande2011}.
HII regions are the most intense sources of recombination line emission, although at low frequencies diffuse ionized gas becomes a relatively bright source of RRL emission \citep[see][]{alve2010, lee2012}.

In the Galaxy, the diffuse ionized gas consists of a low density component ($<$ 1 cm$^{-3}$)
referred to as the warm (T$_{e}$ $\sim$ 3000 to 8000 K) ionized medium (WIM). This component
has a scale height of $\sim$ 1000 pc and is usually studied using optical recombination lines and pulsar dispersion measures 
\citep{tayl1977,reyn1990}. Observations of low-frequency ($<$ a few GHz) RRLs have established the
presence of another diffuse ionized component with density in the range 1 to 10 cm$^{-3}$
close to the Galactic plane and with a scale height of $\sim$ 100 pc \citep[][]{gott1970}. 
In the literature, this component is referred to as Galactic Ridge RRL emission by 
\citet{davi1972},
extended low density medium (ELDM) by \citet{mezg1978}, 
evolved HII region by \citet{shav1976}, HII envelopes
by \citet{lock1976} and \citet{anan1986} \citep[e.g.][]{rosh2000, rosh2001, badd2012}, 
extended low-density warm ionized medium (ELDWIM) by 
Petuchowski \& Bennett (1993) and Heiles (1994)
and Warm ionized medium by Heiles et al. (1996). Following Mezger (1978), 
in the present paper, this diffuse ionized component is referred to as ELDM. 
Recently, extensive higher angular resolution ($14.8\arcmin$) RRL observations
of the ELDM were made by \citet{alve2012}.
They found that the distribution of the ELDM is strongly correlated with the location of 
Galactic HII regions, confirming the observations by \citet{lock1976}, \citet{hart1976}, and \citet{anan1986}.
The origin of the ELDM is, however, still 
unclear, in large part because previous studies have used low-resolution observations.
One possibility is that it originates from the ionization of low density regions surrounding giant HII regions, from which  photons ``leak''\citep[see][for a discussion on 
the W43 region]{ande2011} although this scenario does not fit all observations \citep{rosh2012}.

In addition to hydrogen lines, carbon RRLs have been detected in several directions in the Galaxy.
They are generally observed from interfaces between neutral and fully ionized regions, referred to as PDRs. Photons with wavelengths longer than the Lyman limit can 
escape the HII region and ionize carbon and other atoms of lower ionization potential than hydrogen. Carbon RRLs were first detected 
by \citet{palm1967} and have 
been the focus of many subsequent studies \citep[e.g.][]{pank1977,rosh2011,weng2013}.

\subsection{Motivation for the Survey} 
The Survey of Ionized Gas in the Galaxy, made with Arecibo (SIGGMA) will fully sample the entire Galactic plane observable with the 305-m William E. Gordon Telescope at the Arecibo Observatory ($30 \degr \leq l \leq 75 \degr$ and $ -2 \degr \leq b \leq 2 \degr $ in the inner Galaxy; $175 \degr \leq l \leq 207 \degr $ and $ -2 \degr \leq b \leq 1 \degr$ in the outer Galaxy), and will be the most sensitive large scale RRL survey ever made. 
The survey data will permit a wide range of science, including:
1) studies of HII regions, planetary nebulae, and novae;
2) the Galactic temperature;
3) the large scale structure of the Milky Way;
4) carbon recombination line emitting regions;
and, 5) possibly the ELDM.

RRLs can distinguish between thermal and non-thermal sources. In the section of the inner Galactic plane observable with the Arecibo telescope there are thousands of continuum sources that are revealed by the L-band NRAO VLA Sky Survey (NVSS) \citep{cond1998}. 
To date only a few hundred HII regions have been identified in this zone \citep{lock1989, bani2010}.
With a much greater sensitivity than existing surveys, SIGGMA will detect hydrogen RRLs in sources with peak line intensities $\gtrsim$ 1.5 mJy (3\,$\sigma$ threshold).

HII regions are ideal tracers of spiral arm structure in galaxies.
SIGGMA will offer a large-area sample of Galactic HII regions which, together with the ALFA HI survey of the Galactic plane \citep{peek2011}, will permit a comprehensive study of Galactic structure and kinematics within the Galactic longitude range $30\degr - 75\degr$.
This survey will also be useful to check the current velocity field models of the Galaxy since a rotation curve for the 4.5 to 8~kpc galactocentric distance range can be derived from the data.

\citet {chur1975} found, for the first time, that the average electron
temperatures (T$_{e}$) of
HII regions gradually increase with galactocentric radius R in the Galaxy.
The same trend of T$_{e}$ was also obtained by \citet{shav1983}
who directly related this to the metallicity gradient with R. Using high angular
resolution RRL observations toward Ultra-compact HII (UCHII) regions,
\citet{affl1996}
derived the slope of temperature gradient with R to be 320 K kpc$^{-1}$.
\citet{affl1997} directly determined the metal abundance in these UCHII regions
using IR fine structure line observations. They show, for the first time, that the inferred temperature gradient
from the measured metallicity gradient is consistent with that obtained
using RRL observations.
However, the slope of the electron temperature gradient obtained
from data
towards UCHII regions is shallower than that determined from 
data toward classical HII regions \citep{shav1983, affl1996}.
The difference in slope obtained from the two sets of observations
needs to be resolved.
Moreover, there is still a considerable scatter both in the
temperature and metal abundance gradients.
This scatter is partly due to local metal abundance
anomalies produced by supernovae and
winds from evolved stars, but also due to measurement errors along with uncertainties associated with the estimation of the distance to the objects.
The high sensitivity of SIGGMA will help to improve the estimation of the
electron temperature gradient from the data toward HII regions.

We will also be able to measure RRLs from heavier elements, such as carbon.
Maps of carbon RRL emission in a variety of sources can be used to study 
PDRs and to test PDR models. SIGGMA can make a
greate contribution to the study of PDRs because the Arecibo telescope has 
the sensitivity to map carbon RRLs in a substantial number of sources 
having a wide range of metallicities.

SIGGMA can help to improve our understanding on the origin and ionization of the ELDM since, owing to its high spatial resolution and high sensitivity, it will map the ELDM up to latitudes of $\sim 2\degr$ , thus allowing us to associate it with individual HII regions.


In summary, we expect the data products and results from SIGGMA to be comparable with those obtained from other surveys such as 2MASS \citep[The Two Micron All Sky Survey,][]{skru2006}, ISO \citep[Infrared Space Observatory,][]{kess1996}, MSX \citep[Midcourse Space Experiment,][]{egan2003}, NVSS \citep[NRAO VLA Sky Survey,][]{cond1998}, GLIMPSE \citep[Galactic Legacy Infrared Mid-Plane Survey Extraordinaire,][]{benj2003,chur2009}, IGPS (International Galactic Plane Survey)~\footnote{http://www.ras.ucalgary.ca/IGPS/}, as well as GALFACTS \citep[G-ALFA Continuum Transit Survey,][]{tayl2010} and ALFA Galactic HI Surveys.
\section{The Observations}
\subsection{The Receiver} \label{receiver}
SIGGMA uses the Arecibo L-band Feed Array (ALFA)\footnote{See http://www.naic.edu/alfa for more information} receiver on the Arecibo telescope. The ALFA receiver has 7 independent beams, each recording two orthogonal linear polarizations.
The beams are arranged in a hexagonal pattern such that there is one central beam (Beam 0) surrounded by six outer beams (Beams 1-6).
Due to the optics of the telescope, the outer beams are projected onto an ellipse in the sky. 
This ellipse is centered on Beam 0 and has semi-axes of $6.4\arcmin$ in zenith angle by $5.5\arcmin$ in azimuth. The orientation of the long axis of the ellipse changes with parallactic angle. Although the array is derotated during observations, the positions of the outer beams change on the sky due to the elliptical projection. This change of outer beams with respect to Beam 0 is typically less than the FWHM of the beam width. 

The average FWHM of the beams varies between $3.3\arcmin - 3.4 \arcmin $ across our frequency range\footnote{The source of the average FWHM values as well as of other parameters given in this paragraph is Arecibo Observatory's performance and calibration measures, which are made at 1420 MHz. The errors on the beam area introduced by using the average FWHM and a circular beam rather than the true elliptical beam are at around 1 part per thousand, and thus negligible compared to other sources of uncertainty and to the change in beam size over our frequency range.}.
The footprint of all the beams on the sky, out to the ellipse enclosing the FWHM of the outer beams, is ($15.3\arcmin - 18.6\arcmin) \times (13.6\arcmin - 16.5\arcmin$), covering an area of 163 -- 241 sq. arcmin.
the area falling within the FWHM of the beams is 53 -- 78 sq. arcmin., or around a third of the total foot print. 
The gain of Beams 1-6 is $\sim$ 8.5 K\,Jy$^{-1}$, while Beam 0 has a gain of $\sim 11$ K\,Jy$^{-1}$. 
All seven beams have measured system temperatures of 27 - 28 K at low zenith angles ($\lesssim 15\degr$) where most of our observations will occur, giving a system equivalent flux density of $\sim$ 2.5 Jy for Beam 0 and $\sim$ 3.2 Jy for Beams 1 - 6.
The receiver has a bandwidth of 300 MHz, covering 1225 - 1525 MHz.

\begin{figure}[htbp]
\epsscale{.8}
\plotone{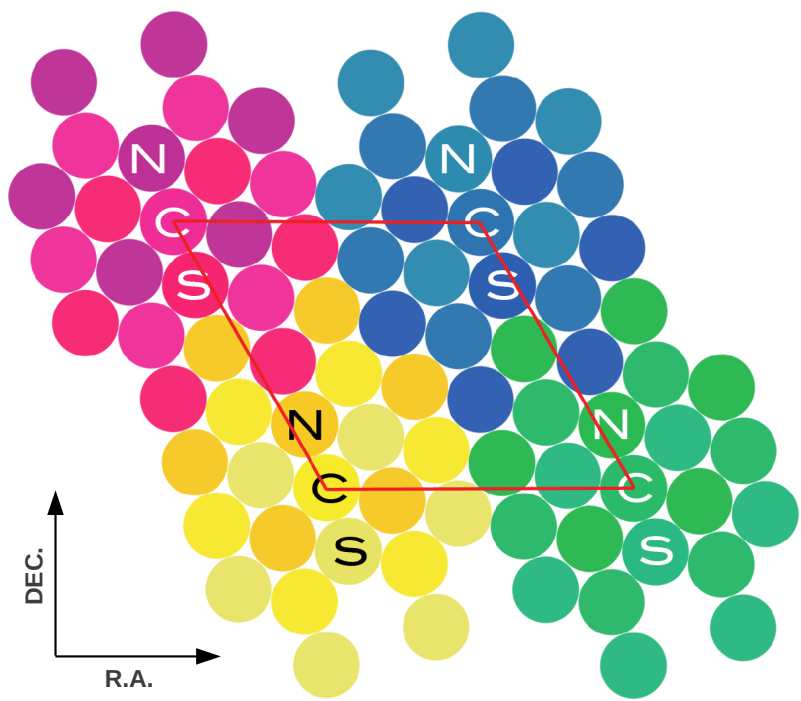}
\caption{The grid of points used for the RRL survey, with RA horizontal and Dec vertical. Circles indicate the size (FWHM) of the ALFA beams. Circles with the same fill color are observed in the same ALFA pointing; circles with similar fill colors are part of the same pointing cluster, with N, C and S indicating the north, central and south pointing of the cluster. The red rhomboid indicates the repeating pattern of the pointing clusters, illustrating that they repeat along lines of constant declination. Developed from Figure 5 of \citet{frei2003}.\label{fig1}}
\end{figure}
\subsection{The Backend} \label{backend}

Spectra are recorded once per second using the Mock Spectrometer, a Fourier-transform device which has 2 groups, each of 14 boards.
This enables it to process data from the 7 ALFA beams, each of which is divided into two IF sub-bands.
The first group is used in a high time-resolution, low spectral-resolution mode to obtain data for the commensal `P-ALFA' pulsar survey \citep{cord2006}, while the second group is used to acquire data simultaneously for the RRL survey.
Each IF sub-band covers 172 MHz, with the first IF centered at 1450 MHz and the second IF centered at 1300 MHz.
Together, these cover the entire 300 MHz ALFA bandpass, with the filter roll-off being either in the overlap region between the two sub-bands, or outside the band of the receiver. 

For the RRL survey, data are accumulated for 1 s before being Fourier transformed to form spectra of 8192 channels per IF sub-band.
This gives a spectral resolution of 21 kHz, equivalent to 4.2 - 5.1 km\,s$^{-1}$ for the Hn$\alpha$ recombination lines within the ALFA bandpass.
The required integration time is then built up from multiple 1 s spectra.

\subsection{Observing Technique} \label{obs tech}

The survey uses a ‘leapfrog’-style observing technique \citep{spit1998} with a grid of points that repeat along lines of constant declination (see Fig.\ref{fig1}).
Each observation integrates on each position for 270 s (180 s in the outer galaxy).
Combined with the slew time, this gives a spacing of nearly 5 min between observations.
By choosing pointings from the grid with the same declinations and separated by 5 minutes in right ascension, we track the same azimuth and zenith angle in consecutive points.
This allows consecutive points to be used as ON-OFF pairs in order to form a bandpass-corrected (ON-OFF)/OFF spectrum (see Section \ref{sec3}).
The 5 minutes separation in time between ON and OFF points provides the best baselines because ripples caused by internal reflections in the antenna, ground pick up, and atmospheric effects mostly cancel out in the corrected spectrum.

The survey tiling pattern is made up of clusters of three pointings that are offset from each other by one beam FWHM beamwidth, labeled as the north (N), central (C) and south (S) pointings in Figure 1.
This is a modification of the original P-ALFA pulsar survey tiling pattern \citep{cord2006}, with the axis of the pattern (and of the ALFA receiver) rotated by 19 degrees to celestial north so that adjacent clusters of pointings fall on lines of constant declination.
By leapfrogging through this tiling pattern, the survey covers approximately 10 sq. deg. of sky in 100~h of telescope time.
In the outer galaxy, where SIGGMA is commensal with the ALFA Zone of Avoidance survey (ALFA ZOA), integration times per pointing are shorter but the pattern is observed three times with a small offset ($\sim 1.9\arcmin$) in order to make a Nyquist-sampled map of 10 sq. deg. of sky to similar depth in 200~h of telescope time.

\section{Data Reduction}\label{sec3}
\begin{figure}[h]
 \epsscale{.65}
 \plotone{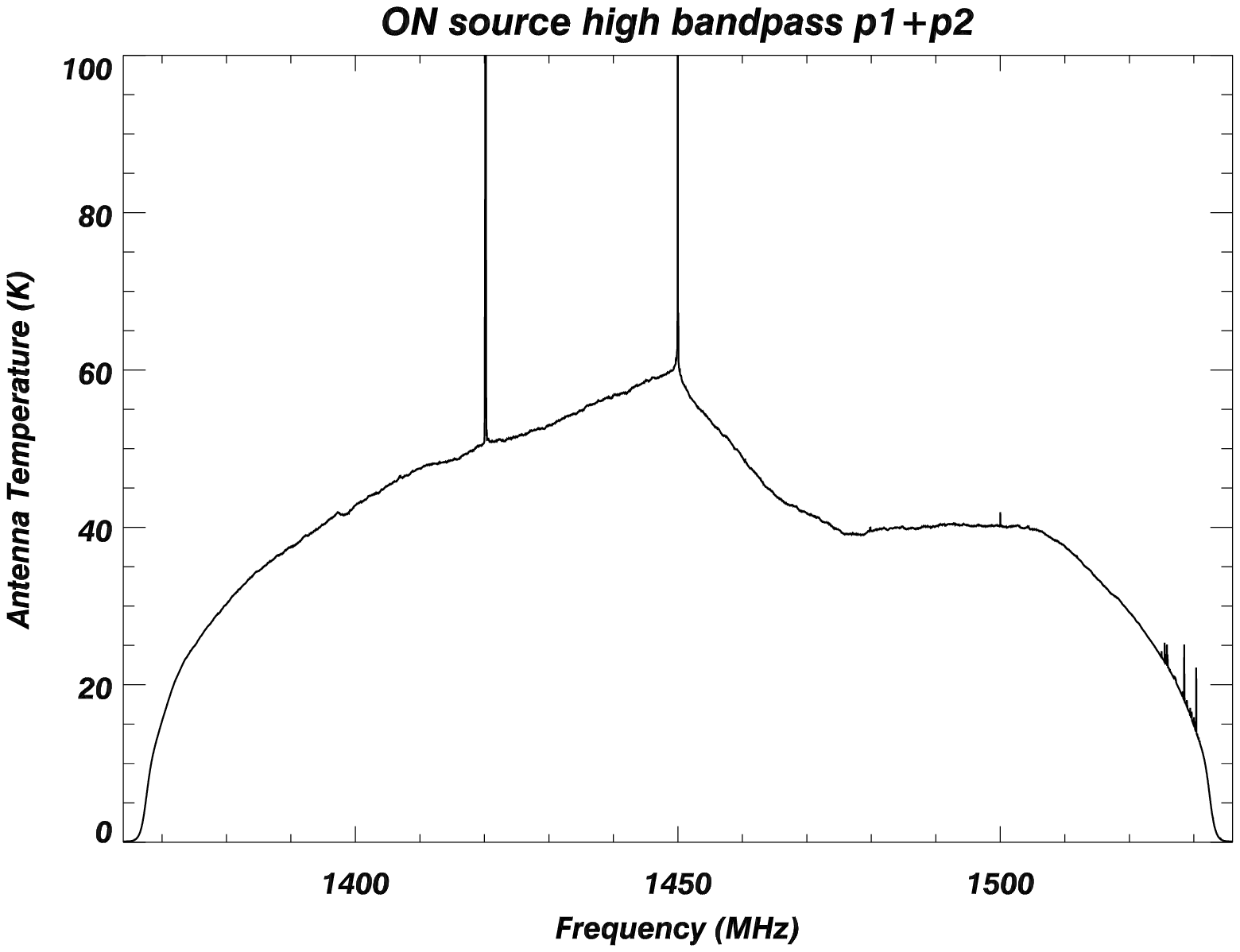}
 \caption{The spectral shape of the raw data for the higher IF sub-band when the antenna beam is on the source S255. This spectrum is produced by averaging over 270 s exposure. The orthogonal polarizations have also been averaged. The antenna temperature includes about 12~K continuum flux from the source and the spikes are due to RFI.\label{fig2}}
\end{figure}

\begin{figure}[htbp]
\epsscale{.65}
\plotone{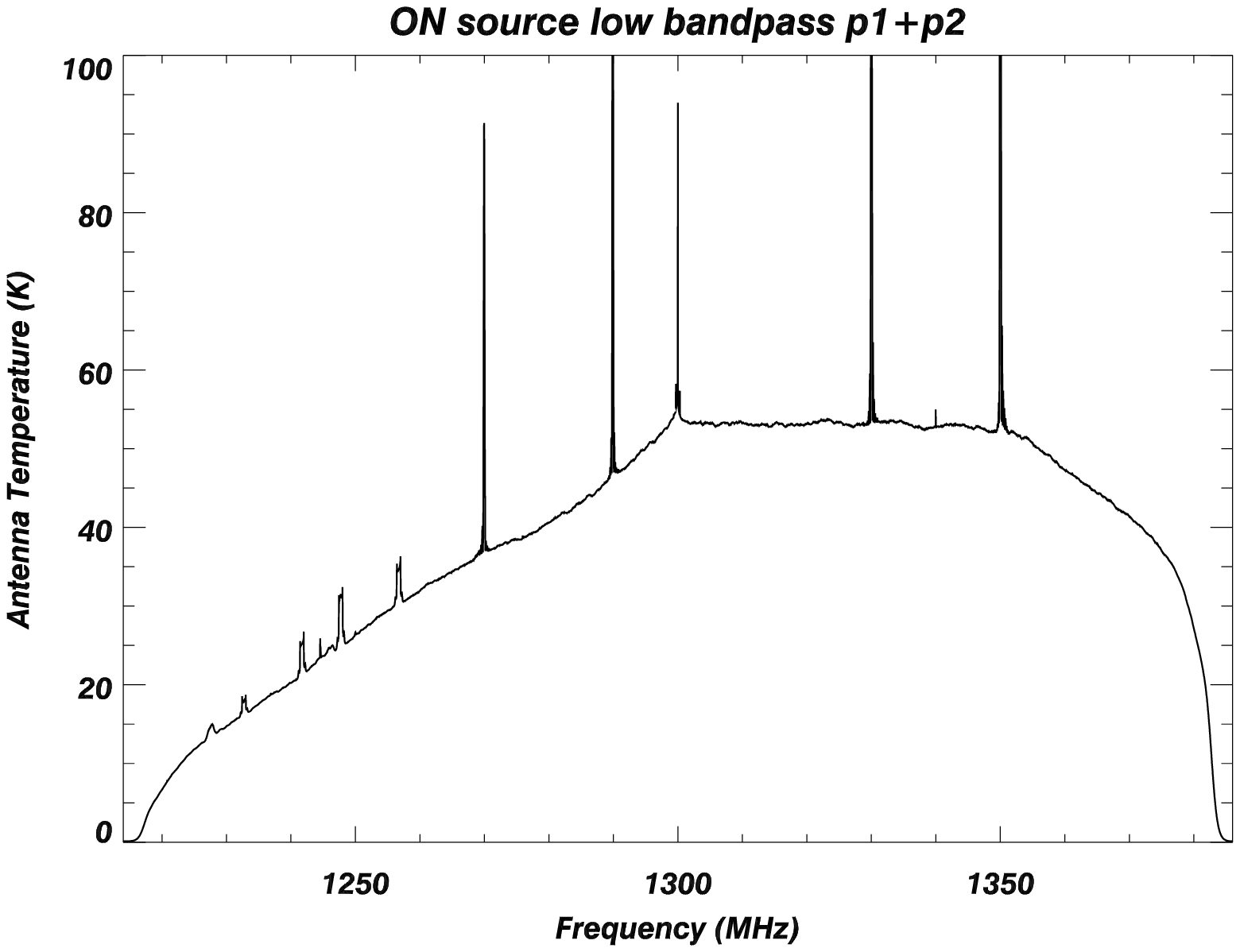}
\caption{Same as Fig. \ref{fig2} but for the lower IF sub-band.\label{fig3}}
\end{figure}

The entire 300 MHz bandpass of the RRL survey observations is divided into two IF sub-bands.
Examples of the spectra of the two sub-bands are shown in Figs. \ref{fig2} and \ref{fig3}.
The two orthogonal polarizations are averaged for each sub-bandpass, thereby achieving $\sim 40 \%$ increase in SN, but rendering the data insensitive to polarization.

\begin{figure}[htbp]
\epsscale{.65}
\plotone{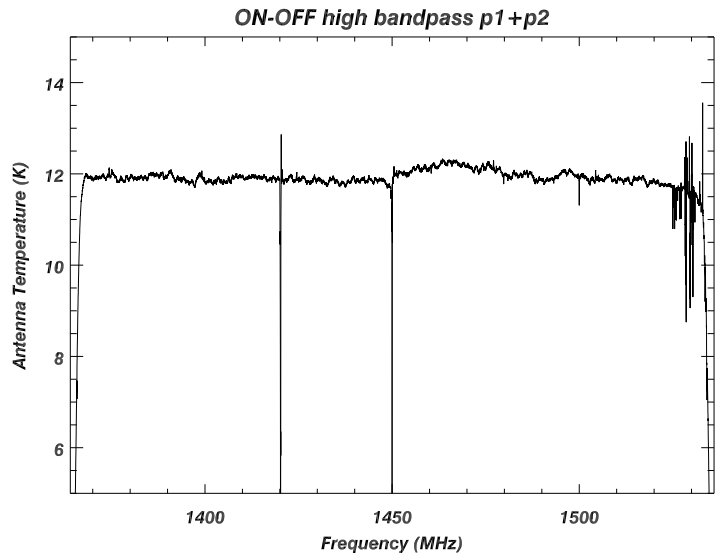}
\caption{The spectrum after (ON$-$OFF)/OFF bandpass correction: the higher IF band. The antenna temperature includes about 12~K continuum flux from the source and the spikes are due to RFI.\label{fig4}}
\end{figure}

\begin{figure}[htbp]
\epsscale{.65}
\plotone{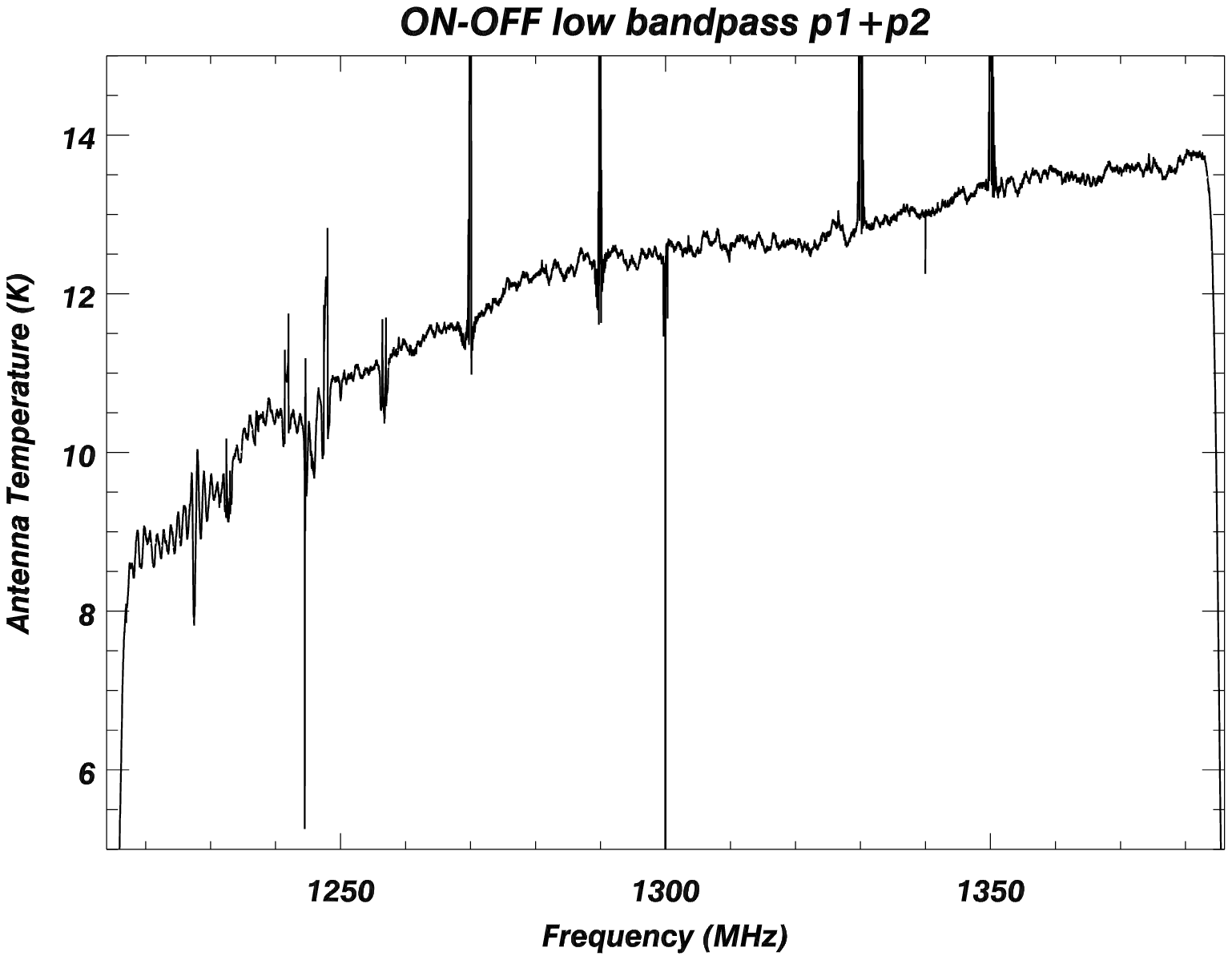}
\caption{Same as Fig. \ref{fig4} but for the lower IF sub-band. \label{fig5}}
\end{figure}

After matching the ON/OFF pointing pairs, we perform bandpass correction to the raw ON/OFF spectral pairs via the customary position switched method of (ON-OFF)/OFF $\times$ Tsys.
Bandpass corrected spectra are shown in Figs. \ref{fig4} and \ref{fig5}.
It is clear that the RFI in the higher frequency IF sub-band is considerably less serious than in the case of the lower IF band. 

We separate each 300 MHz spectrum into 12 segments, each of which covers $\pm$300~km\,s$^{-1}$ centered between the rest frequencies of the Hn$\alpha$ and Hen$\alpha$ lines such that the Hn$\alpha$ and Hen$\alpha$ lines are offset by +61~km\,s$^{-1}$ and $-$61~km\,s$^{-1}$ ,respectively, from the band center and the Cn$\alpha$ line is offset $-$88 km\,s$^{-1}$ from the band center.
The velocity range, with respect to the Local Standard of Rest (LSR), spans the entire velocity of Galactic gas towards each pointing.
Table \ref{tab-1} lists the 12 H, He, and C $\alpha$-transitions located within the full 300 MHz band.


\begin{deluxetable}{ccllll}
\tablecolumns{5}
\tablecaption{The frequencies of the 12 $\alpha$-RRL transitions within the 300 MHz bandpass}
\tablewidth{0pt}
\tablehead{
\colhead{Num} & \colhead{n} & \colhead{Hn$\alpha$} & \colhead{Hen$\alpha$} & \colhead{Cn$\alpha$} & \colhead{Central freq} }
\startdata
1 &163 &1504.608 &1505.221 &1505.359 &1504.9145\\
2 &164 &1477.335 &1477.937 &1478.072 &1477.6360\\
3 &165 &1450.716 &1451.307 &1451.440 &1451.0115\\
4 &166 &1424.734 &1425.314 &1425.444 &1425.0240\\
5 &167 &1399.368 &1399.938 &1400.066 &1399.6530\\
6 &168 &1374.600 &1375.161 &1375.286 &1374.8805\\
7 &169 &1350.414 &1350.964 &1351.088 &1350.6890\\
8 &170 &1326.792 &1327.333 &1327.454 &1327.0625\\
9 &171 &1303.718 &1304.249 &1304.368 &1303.9835\\
10 &172 &1281.175 &1281.697 &1281.815 &1281.4360\\
11 &173 &1259.150 &1259.663 &1259.778 &1259.4065\\
12 &174 &1237.626 &1238.130 &1238.243 &1237.8780\\ 
\enddata
\tablecomments{Col.~1 lists the number of the 12 spectral segments.
Col.~2 gives the 12 lower quantum numbers.
Col.~3-5 are the frequencies of the 12 $\alpha$ lines of H, He and C in MHz.
Col.~6 gives the adopted central frequencies of each narrow band spectrum.}\label{tab-1}

\end{deluxetable}
A 3rd degree polynomial baseline is fitted to each of the 12 spectral line segments.
The 7th and 12th segments are automatically excised, because of bad RFI at these frequencies.
Other segments containing channels with strong RFI are flagged and removed manually.

\section{A Test Observation toward S255}

\begin{figure}[htbp]
\centering
\includegraphics[scale = 0.5]{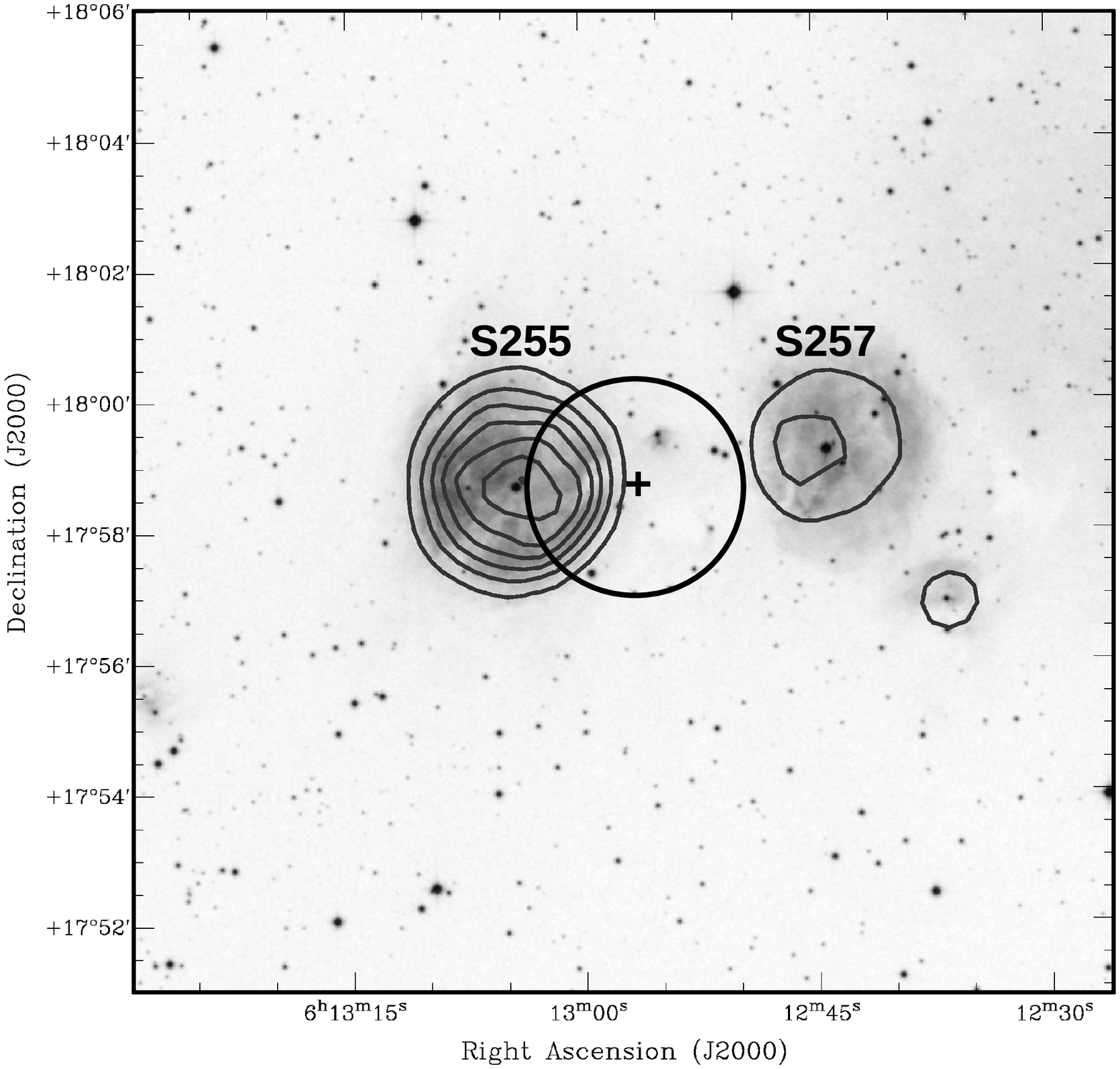}
\caption[]{The map of the S255 - S257 region. It is produced by overlapping the NVSS contours on the optical image from the Digitized Sky Survey (DSS) \footnotemark, which has a $1.7 \arcsec$~pixel$^{-1}$ resolution. 
The plus symbol locates the beam center of our observation and the open circle illustrates the 3.4$\arcmin$ beam size.} \label{s255}
\end{figure}

A test observation for SIGGMA was made toward the HII region S255.
The position (06$^{h}$10$^{m}$01.4$^{s}$, +17$^{d}$59$^{m}$31$^{s}$, B1950) or (06$^{h}$12$^{m}$56.8$^{s}$, +17$^{d}$58$^{m}$40.8$^{s}$, J2000) was observed with Beam 0. We choose this position because RRLs near 10 GHz were observed earlier 
by \citet{lock1989}. However, 
this position is not the center of the optical or radio nebula, but is offset by $\sim 1.5\arcmin$ 
west of S255 and $\sim 3\arcmin$ east of S257 (see Fig. \ref{s255}). In this paper we
refer to this position as S255-west or S255w.

\footnotetext{The Digitized Sky Surveys were produced at the Space Telescope Science Institute under U.S. Government grant NAG W-2166. The images of these surveys are based on photographic data obtained using the Oschin Schmidt Telescope on Palomar Mountain and the UK Schmidt Telescope. The plates were processed into the present compressed digital form with the permission of these institutions. See http://stdatu.stsci.edu/dss/index.html}

\begin{figure}[htbp]
\epsscale{.825}
\plotone{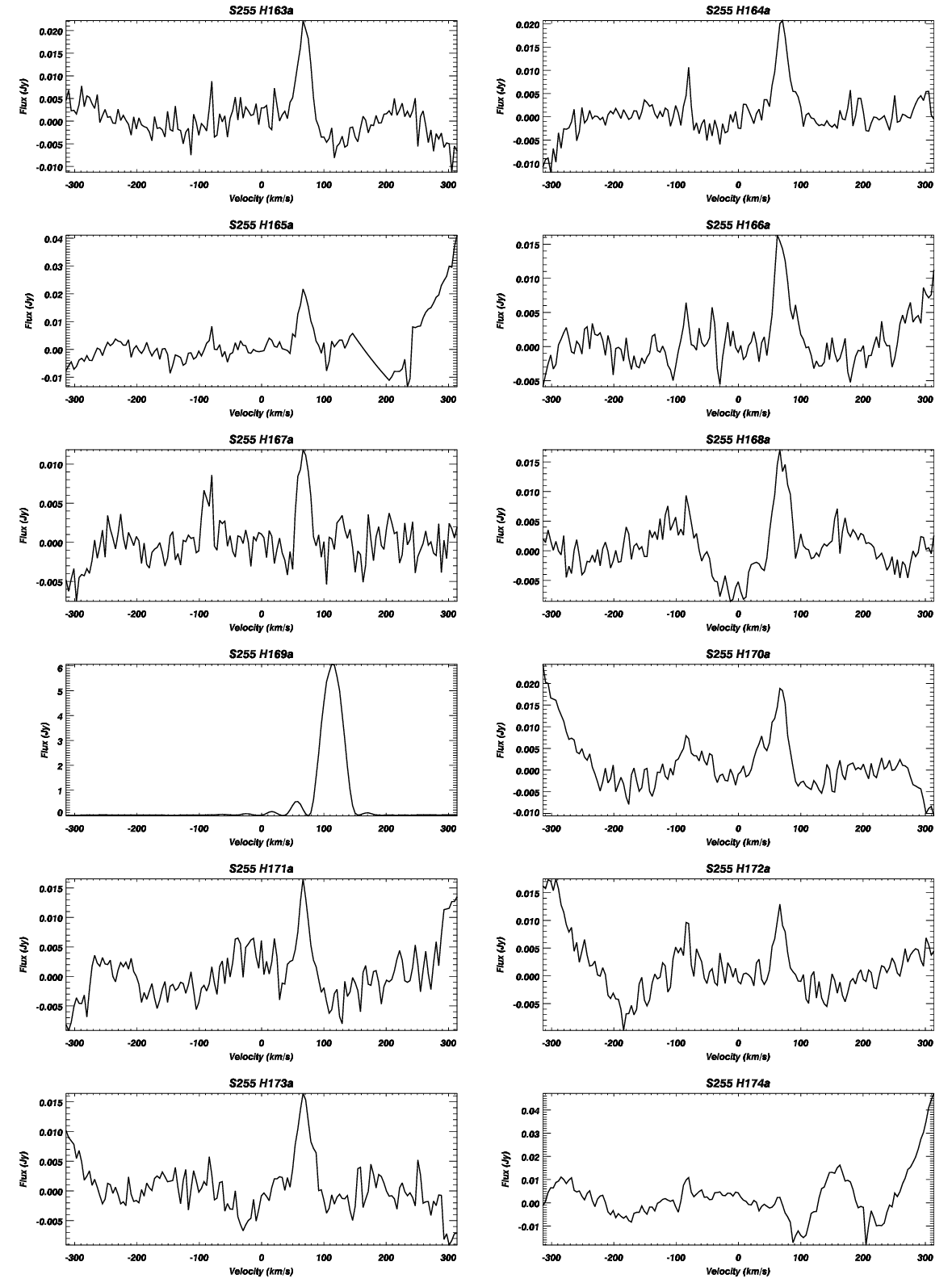}
\caption{The 12 H$n\alpha$ lines in the direction of the HII region S255. The H169$\alpha$ and H174$\alpha$ spectra are spoiled by RFI.
Note that due to the velocity frame set in Section 3, the H$n\alpha$, He$n\alpha$ and C$n\alpha$ lines lie at +61, -61 and -88 km\,s$^{-1}$, respectively, from the origin.}\label{fig7}
\end{figure}

\begin{figure}[htbp]
\epsscale{.8}
\plotone{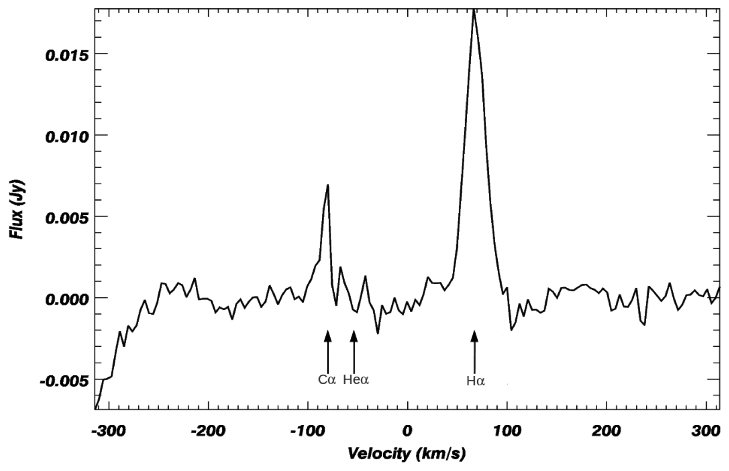}
\caption{The stacked RRL spectrum for the HII region S255.
Note that due to the velocity frame set in Section 3, the H$n\alpha$, He$n\alpha$ and C$n\alpha$ lines lie at +61, -61 and -88 km\,s$^{-1}$, respectively, from the origin.
The drop off on the left-hand edge of spectrum, from $-$300 to $-$250 km\,s$^{-1}$ is caused by the polynomial baseline fitting, and the region has not been included in the calculation of the rms noise on the spectrum.}\label{fig8}
\end{figure}

The twelve lines observed toward S255w are shown in Fig.\ref{fig7}, where the top 6 spectra are from the higher sub-band, and the  bottom 6 are from the lower sub-band.
The Hn$\alpha$ lines can be clearly identified in all the narrow-band spectra that are not 
affected by RFI. Most  of the spectra also show recognizable Cn$\alpha$ lines.
However, strong RFI totally spoiled the  H169$\alpha$ and H174$\alpha$ spectra 
(the 7th and the 12th segments in Fig.\ref{fig7}) rendering them unusable. The
remaining 10 spectra were re-sampled to the same velocity resolution and averaged
to get the final spectrum.
A 5th order baseline was fitted and subtracted from the final spectrum.
Lower order polynomials were tested but they were not sufficient to remove the residual baseline ripple.
The final averaged spectrum for S255w has an rms noise level $\sim$ 0.5 mJy, and is shown in Fig.\ref{fig8}.
A Gaussian fit to the Hn$\alpha$ feature gives a central LSR velocity of +7.1 $\pm$ 0.5 km\,s$^{-1}$, with a FWHM of 23.5 $\pm$ 0.5 km\,s$^{-1}$.
The RRL emission at this position detected by \citet{lock1989} at 10 GHz has a LSR velocity of +7.5 $\pm$ 0.6 km\,s$^{-1}$ with a FWHM 20.1 $\pm$ 1.5 km\,s$^{-1}$, which agrees with our detection.
Parameters obtained for all the line features in the spectrum are listed in Table~\ref{tab2}.

\begin{deluxetable}{clll}
\tablecolumns{4}
\tablecaption{The line parameters of S255w}

\tablewidth{0pt}
\tablehead{
\colhead{} & \colhead{S$_{L}$(mJy)} & \colhead{$v_{LSR}$(km\,s$^{-1}$)} & \colhead{$\Delta v$(km\,s$^{-1}$)}  }\startdata
 H$\alpha$ &15.5$\pm$0.6 &+7.1$\pm$0.5 &23.5$\pm$0.5\\ 
 C$\alpha$ &6.6$\pm$0.7  &+6.4$\pm$0.4 &7.2$\pm$0.5\\
\enddata
\tablecomments{Col.2 is the peak line flux density.
Col.3 gives the line velocity.
Col.4 shows the line width (FWHM).
All the parameters are results of Gaussian fits to the spectrum.
The line widths were corrected for the channel width.}\label{tab2}
\end{deluxetable}

The carbon line detected is almost certainly coming from PDRs.
Since the temperature of these regions is at least a factor of 10 lower than that of the HII region and carbon is ionized in the PDR, intense carbon RRL is expected. 
Carbon RRLs will also be amplified by the background from the HII region (if the PDR is in front of the HII region) and Galactic background.
In order to study the physical properties of these regions observations of carbon RRLs at several frequencies are needed to compare with models which must take the geometry of the region into account.
There is no detection of He RRL in the spectrum of Fig. 8. 
This is in agreement with the results of \citet{silv1979} who did not observe He emission in their sensitive RRL spectrum of S255 at 1.4 GHz. 
For a typical He/H RRL ration of 0.1 \citep{chur1974}, the expected He line strength would be of  about 3$\sigma$.
Therefore the non-detection can be due to the fact that we are not exactly on the source as it can also indicate a lower He/H for this region.

\section{Summary}

The Survey of Ionized Gas in the Galaxy, made with Arecibo (SIGGMA) will be the most sensitive fully-sampled RRL survey of the Galactic plane observable with the Arecibo Telescope. When complete, this survey will cover 300 square degrees: $30 \degr \leq l \leq 75 \degr$ and $ -2 \degr \leq b \leq 2 \degr $ in the inner Galaxy; $175 \degr \leq l \leq 207 \degr $ and $ -2 \degr \leq b \leq 1 \degr$ in the outer Galaxy.
SIGGMA provides fully sampled RRL maps with $3.4 \arcmin$ resolution and a line flux density sensitivity $\sim 0.5$ mJy.
The observations started in 2010 and have covered an area of $\sim 50 $ sq. deg. to date.
A software pipeline has been developed to process and archive SIGGMA data.
The fully-sampled data will also be produced as a set of 3-D data cubes, each of size 
$2 \times 4$ sq. deg. $\times 151$ spectral channels, using the software package, Gridzilla \citep{barn2001}.

Our test observation toward S255/S257 HII complex demonstrates the data quality.
Hydrogen and carbon lines were detected with good signal to noise ratio.
To derive reliable physical parameters such as electron temperature and density as well as metal abundances from SIGGMA data, total continuum data are needed. 
These will be provided by the GALFACTS \citep{tayl2010}, whose observations will be completed in 2013.
We expect to finish observations of regions covering W49 and W51 in the next few months.
These observations will be analyzed using the data reduction pipeline developed and will provide 
early science results for the survey.


\acknowledgments

We thank the anonymous referee for the useful comments on this manuscript.
We are grateful to C. Salter for helpful discussions and comments.
We acknowledge the Arecibo Observatory.
The Arecibo Observatory is operated by SRI International under a cooperative agreement with the National Science Foundation (AST-1100968), and in alliance with Ana G. M\'{e}ndez-Universidad Metropolitana, and the Universities Space Research Association.
B. Liu acknowledges the hospitality enjoyed as a visiting student at Arecibo Observatory and Cornell University.
B. Liu is partly supported by China Ministry of Science and Technology under State Key Development Program for Basic Research (2012CB821800, 2013CB837900) and Projects of International Cooperation and Exchanges NSFC (11261140641). 
\clearpage






\end{document}